# VIRDOCD: a VIRtual DOCtor to Predict Dengue Fatality

[1]Amit K Chattopadhyay and [2]Subhagata Chattopadhyay

[1]Department of Mathematics, Aston University, Birmingham B4 7ET, UK

[2]FAcculi Labs Pvt. Ltd. Bangalore 560098 Karnataka, India

**ABSTRACT**:

Clinicians make routine diagnosis by scrutinizing patients' medical signs and symptoms, a skill popularly referred to as "Clinical Eye". This skill evolves through trial-and-error and improves with time. The success of the therapeutic regime relies largely on the accuracy of interpretation of such sign-symptoms, analyzing which a clinician assesses the severity of the illness. The present study is an attempt to propose a complementary medical front by mathematically modeling the "Clinical Eye" of a **VIR**tual **DOC**tor, using Statistical and Machine Intelligence tools (SMI), to analyze **D**engue epidemic infected patients (*100 case studies with 11 weighted sign-symptoms*). The SMI in VIRDOCD reads medical data and translates these into a vector comprising Multiple Linear Regression (MLR) coefficients to predict infection severity grades of dengue patients that clone the clinician's experience-based assessment. Risk managed through ANOVA, the dengue severity grade prediction accuracy from VIRDOCD is found higher (ca 75%) than conventional clinical practice (ca 71.4%, mean accuracy profile assessed by a team of 10 senior consultants). Free of human errors and capable of deciphering even minute differences from almost identical symptoms (to the Clinical Eye), VIRDOCD is uniquely individualized in its decision-making ability. The algorithm has been validated against Random Forest classification (RF, ca 63%), another regression-based classifier similar to MLR that can be trained through supervised learning. We find that MLR-based VIRDOCD is superior to RF in predicting the grade of Dengue morbidity. VIRDOCD can be further extended to analyze other epidemic infections, such as COVID-19.

## I. INTRODUCTION:

Medical diagnosis is an art of combining supervised learning (as a learner of medical science) with unsupervised/experiential learning (as a practitioner). In both cases, doctors learn why, how and what to look for in analyzing the patterns and inter-relationships between sign-symptom (predictors), as this may vary from case-to-case, patient-to-patient, and even for the same ailment. Doctors then subjectively ascribe weightage to the sign-symptoms diagnosis to understand the morbidity load and its influence on the outcome/diagnosis. Such mental shredding of the symptoms to identify the disease level and nature is an integral component of every diagnostic process.

The numerically translated sign-symptoms are self-narrative that lead to a clinical understanding of the illness in terms of its 'severity' or 'grade'. Such a dynamic medical concept, thus developed within a clinician, is popularly known as the 'Clinical eye' (Shapiro, Rucker, & Beck, 2006). Clinical eye matures as the clinician gains experience. Doctors with matured clinical eyes are able to diagnose a case with reasonably high accuracy, in fact, often pre-empting them (Brunyé, Drew, Weaver, & Elmore, 2019). In the computer science linguistic, this Clinical eye is nothing but a set of conditional statements

---

characterizing each numerical array, e.g., IF-THEN-ELSE serving as algorithmic equivalents of medical sign-symptom 'A', 'B', 'C' further sub-graded by individual weights 'Mild' (M), 'Moderate' (M), and 'Severe' (S) (Lu, Tong, Yu, Xing, Chen, & Shen, 2018), leading to Fatality/Severity/Grade prediction of the disease that are identically characterized within the MMS architecture. A 'specialist' or 'expert' is a clinician who has studied numerous such cases and has thus developed a robust (IF-THEN-ELSE) mental algorithmic map that can relate symptoms to potential severity (also fatality) grade (Chattopadhyay, Banerjee, Rabhi, & Acharya, 2013). An inherent feature of this mental mapping process is the 'subjectivity index' (Alessia Alunno, 2018) whereby different doctors may read symptoms differently, ascribing different weights to the incumbent factors, but eventually converging to a 'global' precision diagnosis that matches other clinicians in the field. Understanding such 'individuality' is critically important in human bias prevention during decision making, especially to avoid undercutting or overdoing the therapeutic regimen (Acharya, Sree, Ng, Chua, & Chattopadhyay, 2014).

Statistical modelling towards Machine Intelligence (SMI) is an evolving domain of Computer Science and Information Technology. It is popularly used for decision making by a Computer that is trained on domain rules to extract causality driven outcomes within the individualized (patient specific) constraints (Chattopadhyay S. , Neurofuzzy Models to Automate the Grading of Old-age Depression, 2014). There are several reasons for the increasing dependence on such SMI assisted clinical decision making tools, some of which are the following: a) as an assistive tool for a second opinion; b) as a nursing aid, to preempt a medical condition before therapeutic intervention; c) as a critical complementary support system, particularly in developing countries, that suffer from acute shortage of medical professionals; d) as a telemedicine tool for beginning medical practitioners; e) as an omnipresent referencing guide, that is ubiquitous in nature. There may be other drivers too. SMI algorithms have already been successfully used in several healthcare domains, such as cardiology (Choi, Park, Ali, & Sungyoung, 2020; Xi he, 2020), mental health (Ashish, Chattopadhyay, Gao, & Hui, 2019), neurology (Dashti & Dashti, 2020), radiology/medical imaging techniques (Jin, et al., 2020; Chattopadhyay, Ray, & Acton, 2005) amongst others. Specialized regression algorithms like pseudo Zernike moment and multinomial regression were successfully used in Alzheimer detection (Wang, et al, 2017) and impending hearing loss (Wang, et al, 2019), including prediction of antimicrobial resistance in ICU-admitted patients (Hernàndez-Carnerero & Sànchez-Marrè, 2021) and (Hernàndez-Carnerero A. , Sànchez-Marrè, Mora Jiménez, Soguero Riuz, Martínez Agüero, & Álvarez Rodríguez, 2020).

The post 2010 era saw a fast emerging landscape of SMI assisted infection modeling (Agrebi & Larbi, 2020) (Silver, et al., 2017). This relates to four key areas – (i) early detection, that can substantially curb morbidity and mortality/case fatality, (ii) early start of treatment typically at the symptomatic stage, (iii) prognostic evaluations, and critically (iv) as a supplement to traditional prognosis tools when they fail to associate events with future prediction of an epidemic due to (a) BIG data size, (b) data complexity, and/or (c) inherent clinical subjectivity (Chen & Asch, 2017). In most cases related to epidemics and pandemics, early detection is of utmost importance in containing infection propagation, thereby reducing the case fatality rate. This is even more pertinent for resource thrifty developing nations, where SMI based tools can provide seamless and ubiquitous healthcare that is hitherto unavailable to the masses (Daneshgar & Chattopadhyay, 2011).

The starting phase of the infection modeling studies relied on conventional Machine Learning (ML) algorithms, typically consisting of k-Nearest Neighbors as part of a supervised learning algorithm (Watkins & Boggess, 2002), followed by creation of *memory kernels* for detecting repeated disease threats (Cuevas, Osuna-Enciso, Zaldivar, Perez-Cisneros, & Sossa, 2012). Support Vector Machine has also been used to accurately detect malarial parasites from RBC (Go, Kim, Byeon, & Lee, 2018). Evolution of infection networks have used optimized Artificial Neural Networks to detect Kyasanur Forst (viral) Disease where the infection load is carried by ticks (Majumdar, Debnath, Sood, & Baishnab, 2018), Ebola propagation severity/outcome (Colubri, Silver, Fradet, Retzepi, Fry, & Sabeti, 2016). The latest addition in the lineage are the Multiple Regression classifiers, e.g. regression algorithm, linear regression model, gradient boosted regression tree algorithm, negative binomial regression model, and generalized additive model, that have shown promise in dengue forecasting in China.

In the recent past, various aspects of Dengue, both epidemiological and clinical pathophysiological, have been studied. Patients' history, sign-symptoms, investigation results are considered as the independent variables, whereas various types of Dengue fevers represent the dependent variables to develop the classifiers. Decision Tree (DT) and Random Forest (RF) classifiers have been used by (Sarma, Hossain, Mittra, Bhuiya, Saha, & Chakma, 2020) to predict Dengue fever. The study concludes that, with 79% accuracy in prediction, DT-based classifier has outperformed RF-based classifier. Tiruveedhula et al (Tiruveedhula, Navya, Gayathri, & Reshma, 2018) applied Simple Classification and Regression Tree (CART), Multilayer Perceptron (MLP), and C4.5 algorithms to analyze the normal and abnormal cases of Dengue using clinical parameters. CART-based classifier performed best with nearly 100% accuracy.

Other algorithms, like ML algorithms, Naïve Baye's, J48, RF, Reduces Error Pruning (REP) Tree, Sequential Minimal Optimization (SMO), Locally Weighted Learning (LWL), AdaboostM1, and ZeroR, have also been used in classifying Dengue data (Rajathi, Kanagaraj, Brahmanambika, & Manjubarkavi, 2018).

Another study targeting early prediction of Dengue incidence in a larger population concluded that the ML-based classifier could detect certain weeks of the year those were found to be vulnerable for dengue outbreak, which would assist the administration and the healthcare setup to get prepared for managing the ailments appropriately. In this study, humidity, wind speed, temperature and rainfall were taken as the independent variables and fed into an SVM classifier whose prediction accuracy, precision, sensitivity, and specificity were found to be 70%, 56%, 14%, and 95%, respectively (Salim, et al., 2021).

In another study with similar objective, i.e., predicting the dengue outbreak timing in an year, authors applied a battery of ML classifiers, e.g., SVM, K-Nearest Neighbor (k-NN), Artificial Neural Network (ANN), Naïve Baye's, DT, Logistic regressions, and LogitBoost ensemble classifier. LogitBoost ensemble classifier was able to predict the outbreak with 92% accuracy (Iqbal & Islam, 2019).

SMI tools have also been used in other areas of data modeling, such as Support Vector Machine (SVM) learning algorithm, Cross-validation (LOOCV) method, and Nested One-versus-one (OVO) SVM. The latter was used to analyze gene sequences from bacteria in preference to the high-resolution melt (HRM) method. The combination of SVM and HRM has been shown to identify bacterial colonies (Fraley, et al., 2016) with high accuracy (100%). SMI based epidemiological models are known to successfully complement error ridden laboratory procedures relating to sample collection, preservation, distribution, and laboratory testing, e.g. assessing fatality due to pulmonary Tuberculosis, the second most frequent cause of deaths (Saybani, et al., 2015), especially of the multi-drug-resistant variety (Huddar, Svadzian, Nafade, Satyanarayana, & Pai, 2020), or Cardiovascular (CVD) risk with lifestyle changes (Xi He, et al 2020). A topic that is assuming critical importance during the present Covid onslaught is the SMI interpretation of herd immunity (O'Driscoll, et al., 2020), especially in predicting its emergence (Chattopadhyay, et al 2021).

The state-of-the-art literature clearly points to three important knowledge gaps:

(i)     None of the ML-based classifiers used in these analyzes integrate the rule-bases of the human clinicians with those from the model, thus making these studies less robust clinically.

(ii)    The earlier studies use MLR, RF and other classifiers to classify the data points through a form of unsupervised learning. VIRDOCD conceptualizes MLR and RF-based classifiers as 'learning tools', based on its coefficient values, entropies, as well as Gini index, to analyze data modeled outcomes through the lenses of seasoned clinicians.

(iii)   Many of these studies lack cross-validation against other classifiers, unlike in this study.

Structured on these three research questions, the key deliverable of this study is a tool that can easily integrate with a medical setup that is usable both by clinicians and nurses, thereby, doubling up as a *Virtual Doctor* (VIRDOCD). The aim is not to substitute or even downplay the role of human intervention but rather to serve as a complementary diagnostic aid. A key technical novelty of this study is the reinvention of *intelligent statistical modeling* as an equally powerful diagnostic tool, substituting the more popular choice of deep learning algorithms that are more complex and hence difficult to maneuver. VIRDOCD can be a layman's tool, that is self-contained, and with attributes that can be sourced in individualized healthcare.

Section II of the article outlines the Experimental design; section III illustrates the results obtained; section IV summarizes the conclusions from this virtual model and highlights on future extensions.

## II. EXPERIMENTAL DESIGN:

The numerical experiment uses a 6-stage data modeling architecture that is divided into **(A)** Data collection from various sources taking proper ethical measures (Chattopadhyay S. , 2012), **(B)** Data preprocessing and fidelity check (Goforth, 2015), **(C)** Data mining – examining within group and between group variations of the collected data by 1-WAY ANOVA, (Anwla, 2020) **(D)** Development of predictive model using Multiple Linear Regressions (MLR) (Rao, 2020), **(E)** Testing the model performance on a set of test cases where outputs are known, **(F)** Parametric study to observe how each of the individual input parameters influences the prediction, as well as their cross-correlated cumulative contribution towards the prediction performance of the model, and **(G)** Comparing MLR-based classifier's performance accuracy with a Random Forest (RF)-based classifier and then validating against human clinicians.

### A. Data Collection:

Primary data (N=100) were enumerated from bed tickets and prescriptions. Data collection processes and activities are outlined in **Table 1** below.

**Table 1. Data Collection parameters**

| | | |
|---|---|---|
| ***Duration***: Jan 2018 – Dec 2019 (Two years) | ***Source***: Clinics, Nursing homes in the vicinity and hospitals | ***Diagnosis***: confirmed with NS1 rapid test and Elisa IgM and IgG |
| ***Habit of substance abuse and alcoholism***: ignored | ***Patient population***: 150 | ***Gender***: Males - 98, Females – 52 |
| ***Age group***: 18 yrs and | ***Co-morbidities***: ignored | ***Socio-economic-condition***: |

| above | | ignored |
|---|---|---|
| **Ethical measures and data privacy protection**: Data source, doctors' and patient names and address/telephone numbers remain anonymous. | | |

**Clinical 'input' parameters**: A total of 11 "sign-symptom" (Sahak, 2020), as follows,

- Fever **(F)**
- Vomiting **(V)**
- Diarrhea **(D)**

- Sore throat **(S)**
- Stomachache **(ST)**
- Joint pain **(J)**

- Headache **(H)**
- Myalgia **(M)**
- Bleeding gums **(B)**

- Nausea **(N)**
- Rashes **(R)**

**Conventional sign-symptoms Map for Symptomatic Dengue Analysis:**

1. *Fever*: This is the most common symptom in symptomatic dengue cases, sometimes exacerbated due to viral load in blood (viraemia).
2. *Sore throat*: Due to involvement of the upper respiratory tract.
3. *Headache*: Due to the associated sinusitis as the consequence of upper respiratory tract infection.
4. *Nausea*: Due to viraemia.
5. *Vomiting*: Due to viraemia
6. *Stomach ache*: Due to bleeding in the rectus muscle sheath.
7. *Myalgia*: Due to diffused viral invasion in muscles causing inflammation.
8. *Rashes*: Due to capillary dilatation under the skin.
9. *Diarrhea*: Due to excessive fluid generation inside bowel.
10. *Joint pain*: Due to inflammation of the joint.
11. *Bleeding gum*: Due to lowering of platelet count.

**Weighted values assigned to each clinical 'input' parameter:**

*Fever:* Measured by thermal scanner; 3-point scale – *Mild* (99º F < m < 101º F), *Moderate* (101º F < M < 102º F), *Severe* (S > 102º F).

*Sore throat:* Patient reported, clinically tested; cumulative weights *w [0, 1]* ascribed on a 3-point scale – *Mild* (w ≤ 0.33), *Moderate* (w ≤ 0.66) and *Severe* (w > 0.66).

The remaining parameters are similarly assessed over a 3-point mMS scale.

**Clinical values of the 'outcome' parameters**: As noted, we classify disease outcomes (O) in 3 categories – mild (m), moderate (M) and severe (S). Weighted scores [0, 1] are drawn from patients' reports and feedback from clinicians. These values are then subdivided into a 3-point outcome-scale as follows:

*Mild (m) (O ≤ 0.33):* patients have responded to symptomatic home treatment amounting to full recovery.

*Moderate (M) (0.33≤O≤0.66):* systemic complications in patients, leading to hospital treatment, eventual therapeutic management ensuring full recovery.

*Severe (S) (O>0.66):* patients had to be shifted to ICU/ITU amounting to increased recovery time or fatality. In this study, patients with O *0.66* became critically ill, but none expired.

Together, this 3-point classification of severity calibration is defined as 'mMS'. Note that the cut-off values used (0.33 and 0.66) relate to one-third and two-thirds number density of cases; different cut-off markers could also be subjectively implemented.

The statistical modeling and predictive ML algorithm in this study were implemented through Python, set within the panda, matplotlib, scipy, numpy, math and sklearn environments (data and code to be released through open access repositories).

## B. Data pre-processing and fidelity check:

Input data, presented as csv spreadsheet, comprise clinical parameters recorded from inputs by attending clinicians. The result was expressed as a 3-dimensional (N X P x K) asymmetric matrix, where '$N$' (=100) denotes the number of cases/patients, '$P$'(=11) the clinical parameters and '$K$' (=3) refers to the corresponding 3-point outcome possibilities (mild, moderate, severe). The operator matrix is thus represented as follows

$$N_1 \, x \, M_j \to K_k \qquad\qquad (1)$$

Data "x" thus collected were (column) normalized between [0,1] using 'Max-Min normalization' method (McCaffrey, 2020), leading to a min-shifted data set normalized within the maximum-minimum values:

$$y = (x - min) / (max - min), \qquad\qquad (2)$$

where '$min$' is the minimum cell value and '$max$' is the maximum cell value correspond to parameters '$P$' and '$K$', as defined in Eq (1). This technique linearly maps the variable '$x$' to '$y$' in a continuous number space varying between 0 and 1 without any data loss, which is a significant advantage. Note, our choice of 'min' value is one that is close to the baseline '0' but not exactly at '0' while '$max$' approaches '1' but is not exactly at '1'. The uncertainty windows around the two limiting values account for subjectivity in diagnosis that are known to fluctuate both with patients and clinicians alike.

Parameters '$P_i$' (i = 1, 2 …, 11) and outcomes '$K_j$' (j = 1, 2, 3) follow the same 3-point mMS scale as before - '$mild$' (m<0.33), '$moderate$' (0.34<M<0.66) and '$severe$' (0.66<S<1.0, marked by the cell color red).. Table 2 depicts a representative data set of the mMS responses.

Representative pre-processed (normalized) data of the 100 dengue cases are shown in **Table. 2**. The VIRDOCD algorithm was trained to adapt to the 3-point weighted sign-symptoms based dengue markers as advised by the clinicians. The target was to develop a VIRtual DOCtor through supervised learning that can self-sufficiently ascribe severity scores to patients, independent of clinicians, thus ensuring 'independent' unbiased decision making rid of human subjectivity errors in diagnosis.

**Table 2. Glimpse of a set of pre-processed data after 'column-wise' max-min normalization**

| F | ST | H | N | V | S | M | R | D | J | B | O |
|---|----|---|---|---|---|---|---|---|---|---|---|
| 0.4235 | 0.9951 | 0.7885 | 0.7903 | 0.9890 | 0.0129 | 0.5451 | 0.4906 | 0.5829 | 0.2238 | 0.2473 | 0.9501 |
| 0.0168 | 0.2689 | 0.1986 | 0.8332 | 0.6194 | 0.3411 | 0.0028 | 0.7765 | 0.3140 | 0.1712 | 0.4704 | 0.5430 |
| 0.2407 | 0.0060 | 0.9032 | 0.0126 | 0.7929 | 0.2719 | 0.8087 | 0.3304 | 0.8788 | 0.7399 | 0.0764 | 0.0910 |
| 0.6419 | 0.7956 | 0.7982 | 0.1080 | 0.3675 | 0.5669 | 0.7496 | 0.2794 | 0.9861 | 0.1896 | 0.5005 | 0.3661 |
| 0.1579 | 0.8692 | 0.3671 | 0.2629 | 0.2402 | 0.5601 | 0.3382 | 0.3035 | 0.6949 | 0.2579 | 0.9715 | 0.6669 |
| 0.6255 | 0.2409 | 0.1061 | 0.9372 | 0.4081 | 0.0943 | 0.8603 | 0.3632 | 0.3883 | 0.2331 | 0.7141 | 0.6668 |
| 0.7801 | 0.4869 | 0.0794 | 0.1635 | 0.2083 | 0.0769 | 0.9368 | 0.8339 | 0.1505 | 0.4656 | 0.5534 | 0.3540 |
| 0.2068 | 0.5325 | 0.7005 | 0.2269 | 0.5064 | 0.7905 | 0.1602 | 0.7341 | 0.0556 | 0.0105 | 0.6994 | 0.6268 |
| 0.4490 | 0.3391 | 0.1189 | 0.9998 | 0.7863 | 0.4638 | 0.7706 | 0.0666 | 0.7818 | 0.3005 | 0.4936 | 0.5507 |

**Data reliability testing with Cronbach's α:**

To establish eligibility/fidelity, that is how closely related the 11 variables are in defining the dengue infected group and analyze their interconnectedness, the data were pre-processed/mined using Cronbach's $\alpha$ method (Cronbach, 1951) (Goforth, 2015):

$$\alpha = \text{¿¿} \tag{3}$$

Here '$r$' refers to the number of scaled data, $\bar{c}$ is the mean of all covariances between data points, and $\bar{v}$ is the average variance. The Cronbach measure checks for internal consistency of the dataset and is the most important pre-processing step. The consistency score α is expressed as a number between 0 and 1, where α$\geq$0.8 is considered ideal while α$\leq$0.5 is deemed "unacceptable" (Goforth, 2015).

## C. Data mining:

**Descriptive statistics:**

This is a process where epidemiological data is expressed as a functional combination of its features and quantifying parameters like shape, frequency, central tendency (mean, median represented by 50% percentile, and mode), dispersion (range, standard deviation, variance), and position (percentile rankings, quartile ranking) (Sucky, 2020).

**Analysis of Variance (ANOVA):**

Analysis of the Variance Test (ANOVA) is a generalization of the t-tests involving more than two groups (Fisher, 1921). ANOVA quantifies the difference in the mean value anywhere in the model (checking for a 'global' effect), but without informing where the difference lies (if there is any). To find where the difference is in between groups, post-hoc tests are required (Anwla, 2020). One-way ANOVA has been conducted in this work to examine whether the sign-symptoms, classed under mMS categorization (variable "K"), are statistically different from each other. It has also been conducted to understand whether the clinical grades (fatality/grade/severity) significantly differ between (dengue) patients. For cases with statistically significant outcomes from one-way ANOVA, the Alternative Hypothesis ($H_A$) was used instead of the Null Hypothesis ($H_0$), indicating that there were at least two groups which are statistically significant while being different from each other (Anwla, 2020). The Null Hypothesis was validated against F-statistics, the ratio of variance of the group means to mean of the intra-group variances. F=1 points to null hypothesis. It tests the null hypothesis using equation 4, below,

$$H_0 = \mu_1 = \mu_2 = \ldots = \mu_k. \tag{4}$$

Here $\mu$ represents the mean of the group and $k$ measures the number of such groups. If, however, the one-way ANOVA returns a statistically significant result, the Alternative Hypothesis ($H_A$) is accepted instead of the Null Hypothesis ($H_0$), indicating that there are at least two group which are statistically significantly while being different from each other (Anwla, 2020).

F-statistic (the ratio of variance of the group means / mean of the within group variances) equals to 'one' accepts null hypothesis, else it accepts alternative hypothesis. For this work, F-static values (Anwla, 2020) are shown in the Results section.

**D.  Development of VIRDOCD with Multiple Linear Regressions (MLR):**

Multiple Linear Regression (MLR) is an extension of the linear regression model (Rao, 2020) that combines multiple explanatory variables to predict one or more response functions. VIRDOCD uses MLR supervised learning strategy, trained by the 'Clinical eyes' i.e., the knowledge base accorded by a set of experienced human doctors. Unlike human intervention, a key motivation behind this strategy is to outline a diagnostic regime that can stay unperturbed (i.e., the retention of 'individuality' in clinical decision making) to changes in rule/knowledge bases according to the judgments of individual clinicians. If $Y$ represents the predicted output i.e. the weighted Case fatality grade of a set of N patients with sign-symptoms $X_i$, ($i =1, 2, ..,11$) weighted by w[0,1]), the working principle of the MLR model is given by

$$Y = B_0 + B_1 X_1 + B_2 X_2 + \ldots + B_p X_p, \tag{5}$$

where the $B$'s are the slope coefficients for individual predictors. The entries for variables $X_i$ are obtained from hospital inventory, a represented set of which is shown in Table 2. The aim is to find the best-fit $B$ values that minimize the error functions extrapolating the best line or hyperplane depending on the number of input variables/predictors (weighted sign-symptoms). Null hypothesis is accepted when all coefficients ($B$) are zero, else we accept alternative hypothesis, that is, when at least one $B$ value is non-zero, that amounts to linearly independent variables. Predicted degree of dengue fatality using the Coefficient values ($B$) and 11 sign-symptoms ($X$ values) for few cases are described in the Results section.

**E. Testing the performance of VIRDOCD:**

VIRDOCD was trained on 75% of the dengue data i.e. 75 patients, mimicking the 'Clinical Eyes' or 'Rule base' (represented as the IF-THEN rules) of ten experienced clinicians. The model was then tested for its predictive power by using the remaining 25% and comparing the predictions against medical prognosis (by the same medical team). 'Ten-fold cross validation' was conducted by randomly partitioning the dataset into 10 equal sized compartments. Of the 10 compartments, one was retained for data validation and associated testing of the model; the remaining 9 compartments were used for training data (de Rooij & Weeda, 2020).

Prediction errors were measured using the Root Mean Square (RMS) formula ($\sqrt{\dfrac{1}{N}\sum_i X_i^2}$). RMS Error (RMSE) was preferred over Mean Square Error (MSE) as RMSE accord relatively higher weights to large errors. This means RMSE is a more accurate measure of fluctuation when large errors are undesirable, which is expected and desired in clinical decision making (Willmott & Matsuura, 2005). RMSE thus obtained has been shown in Table 3.

**F. Parametric study to check the sensitivity (robustness) of the model:**

VIRDOCD was calibrated for robustness using a 2-stage strategy:

(a) Single-factor influence, where predictions by VIRDOCD and other doctors were compared by varying the weights of only one predictor (sign-symptom) for any given dengue case.

(b) Cumulative influence, where comparisons have been made by varying weights of all the sign-symptoms for any given dengue case.

**G. Validation of VIRDOCD:**

VIRDOCD's performance in predicting Dengue is validated using a two-stage approach,

(a) Comparing its performance with an *'RF-based classifier'* developed in this study, working principle of which is detailed below.

(b) Comparing with the *human clinicians'* accuracy of prediction.

**Random Forest (RF) regression:**

RF is a bundle of Decision Trees (DT) that is created randomly. It is also a regression-based classifier, which learns using a supervised learning method similar to MLR. Hence, RF has been chosen for comparing VIRDOCD's performance. Merging of individual DT's output into one gives the final output of the algorithm. RF's training is conducted by the 'bagging' method, where combinations of learning yield the final result (Donges, 2019). **Figure 1** shows the workflow diagram of the RF algorithm. RF can do regressions and classification simultaneously similar to MLR. RF is also a useful algorithm to handle multi-dimensional data and prevent data over-fitting as MLR is capable of (Jaiswal, 2021).

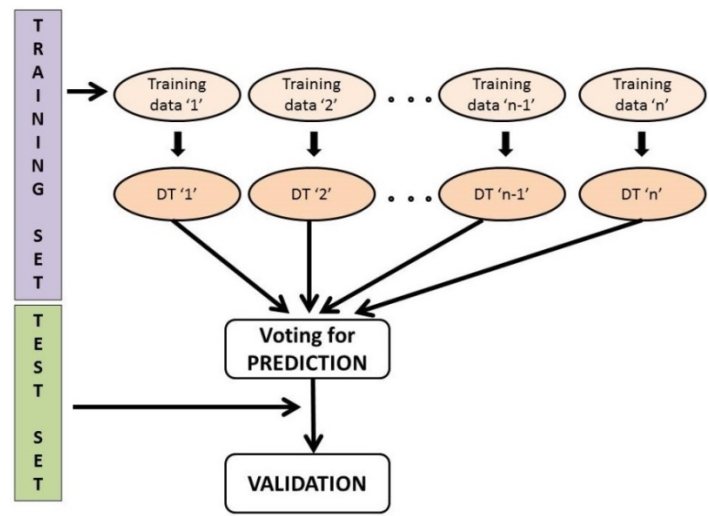

**Figure 1. Working principle of a RF regression algorithm.**

The working steps of an RF algorithm are as follows:

Step-1: choosing random 'k' data points from the Training set

Step-2: building DT subsets of each

Step-3: choosing 'N' number of DTs (here, we have chosen '3', for Mild, Moderate, and Severe grade of sign-symptoms)

Step-4: repeat Step-1 to 3

Step-5: for each test data point, finding predicted values of each tree

Step-6: assigning the test data points to the category that wins the maximum vote.

The RF-regression is computed by estimating the MSE using the following Eqn. (6), where $p_i$ and $t_i$ indicate the predicted and target outputs respectively, such that $1<i<N$.

$$MSE = \frac{1}{N} \sum_{i=1}^{N} \left( p_i - t_i \right)^2 \dots (6)$$

Classification using RF can be made by calculating the Gini-index ($GI$) as in Eqn. (7) below, where $f_i$ represents the relative frequency of a class and '$c$' the number of such classes.

$$GI = 1 - \sum_{j=1}^{c} f_j^2 \dots (7)$$

Entropy (E) has been measured to analyze nodal branching in the DT. This follows Eqn. (8) below:

$$E = \sum_{i=1}^{c} -f_i . \log_2 \left( f_i \right) \dots (8)$$

The results of all experiments are outlined in the following section.

## III. EXPERIMENTAL DESIGN: RESULTS AND DISCUSSIONS

In this section the results of the experiments have been showcased with necessary analysis of the results.

**A. Cronbach's α:** A score of *0.8311* indicates that the data is internally consistent. **Table 3** below shows parameter-wise (sign-symptom-wise) representation of the evenly distributed medical data (refer to boxplots in **Figure 2** for visualization) without noticeable difference between the minimum (min), maximum (max), mean and standard deviations (std in the table). Similar results were obtained in the quartile distributions. Boxplots represent the five-number summary of minimum, first quartile, median, third quartile, and maximum.

**Descriptive statistics:** below Table 3 shows the descriptive statistics of all parameters under test.

**Table 3. Descriptive statistics**

|      | F | S | H | N | V | ST | M | R | D | J | B | O |
|------|---|---|---|---|---|----|----|---|---|---|---|---|
| count | 99 | 99 | 99 | 99 | 99 | 99 | 99 | 99 | 99 | 99 | 99 | 99 |
| mean | 0.4638 | 0.4833 | 0.5455 | 0.4986 | 0.5173 | 0.5310 | 0.5689 | 0.4807 | 0.4838 | 0.4760 | 0.5190 | 0.5009 |
| std | 0.2968 | 0.2761 | 0.2826 | 0.2830 | 0.3110 | 0.3176 | 0.2835 | 0.2849 | 0.2915 | 0.2913 | 0.2878 | 0.2867 |
| min | 0.0011 | 0.0060 | 0.0189 | 0.0010 | 0.0113 | 0.0067 | 0.0028 | 0.0022 | 0.0107 | 0.0105 | 0.0137 | 0.0039 |
| 25% | 0.1941 | 0.2569 | 0.3558 | 0.2631 | 0.2498 | 0.2625 | 0.3383 | 0.2925 | 0.2539 | 0.2218 | 0.2564 | 0.2567 |
| 50% | 0.4712 | 0.4869 | 0.5697 | 0.5416 | 0.5184 | 0.5601 | 0.5451 | 0.4807 | 0.4455 | 0.4497 | 0.5391 | 0.5154 |
| 75% | 0.7188 | 0.7349 | 0.8056 | 0.7419 | 0.8051 | 0.8108 | 0.8165 | 0.7317 | 0.7255 | 0.7484 | 0.7596 | 0.7303 |
| max | 0.9650 | 0.9951 | 0.9817 | 0.9998 | 0.9976 | 0.9978 | 0.9908 | 0.9804 | 0.9999 | 0.9913 | 0.9829 | 0.9890 |

**Table 3** presents sign-symptoms (parameters) data with min-max-std fluctuations (refer to boxplots in **Figure 2** for improved visualization) without much difference between the minimum (min), maximum (max), mean and standard deviations (std, in the table). Similar results were obtained in the quartile distributions. Boxplots are the five-number summary of minimum, first quartile, median, third quartile, and maximum. **Figure 2** shows horizontal lines varying between [1,0], in a top-bottom orientation where the horizontal line at the bottom denotes minimum while the uppermost represents the maximum. The weighted (human) knowledge base from **Table 3**, comprising inputs from ten senior clinicians, can serve as good training material VIRDOCD.

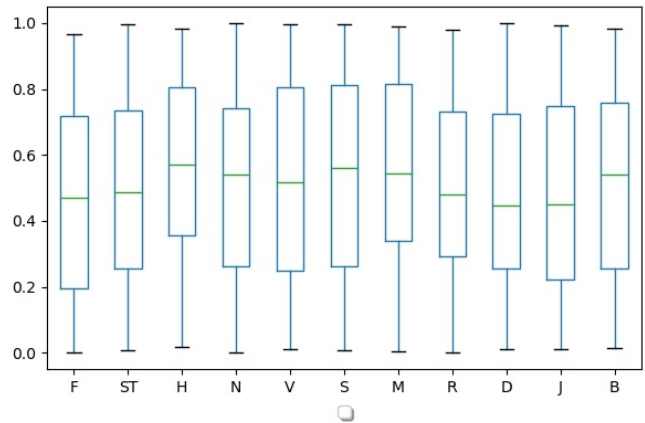

Figure 2. Boxplots of input parameters (sign-symptoms).

### DATA MINING: ONE-WAY ANOVA:

Analysis of Variance (ANOVA) has been used to determine whether there are any statistically significant differences between the means of two or more independent (unrelated) sign-symptoms as in the 3-point scale (mild, moderate, and severe) discussed before.

**Table 4. ANOVA table**

**Table 4** represent residuals (experimental errors) that are *normally* distributed. The probability plot of model residuals (Shapiro-Wilks Test) (Das & Rahmatullah Imon, 2016), that is sign-symptoms weighted between [0,1] is given in **Figure 3**. Observations are sampled independently of each other. In this experiment, the F-statistic was found unequal to 1 for any of the parameters, thus refuting the null hypothesis and accepting alternative hypothesis. If the p-value is less than $\alpha\lfloor 0.95\, here\rfloor$, the null hypothesis is rejected. **Table 4** clearly shows all p-values (predictor-wise) under 0.95, and hence they affect the outcome and accept alternative hypothesis, rejecting null hypothesis. Also note that since p>0.05, the sign-symptoms profile follow a non-Gaussian probability density function (**Figure 3**). **Table 4** provides prediction for a given (first) dengue case conducted to test the fidelity of the code and the mathematical formula using Eq. (5).

|  | sum_sq | df | F | PR(>F) |
|---|---|---|---|---|
| F | 0.015345 | 1 | 0.185085 | 0.667993 |
| Residual | 8.042192 | 97 | NaN | NaN |
| ST | 0.218884 | 1 | 2.708591 | 0.103046 |
| Residual | 7.838654 | 97 | NaN | NaN |
| H | 0.067007 | 1 | 0.813418 | 0.369346 |
| Residual | 7.990531 | 97 | NaN | NaN |
| N | 0.313501 | 1 | 3.926842 | 0.050352 |
| Residual | 7.744036 | 97 | NaN | NaN |
| V | 0.001303 | 1 | 0.015687 | 0.900587 |
| Residual | 8.056235 | 97 | NaN | NaN |
| S | 0.162113 | 1 | 1.99166 | 0.161367 |
| Residual | 7.895424 | 97 | NaN | NaN |
| M | 0.005418 | 1 | 0.065267 | 0.798897 |
| Residual | 8.052119 | 97 | NaN | NaN |
| R | 0.10311 | 1 | 1.257372 | 0.264917 |
| Residual | 7.954427 | 97 | NaN | NaN |
| D | 0.001197 | 1 | 0.014407 | 0.904709 |
| Residual | 8.056341 | 97 | NaN | NaN |
| J | 0.032043 | 1 | 0.387286 | 0.535189 |
| Residual | 8.025494 | 97 | NaN | NaN |
| B | 0.024667 | 1 | 0.297862 | 0.586479 |
| Residual | 8.032871 | 97 | NaN | NaN |

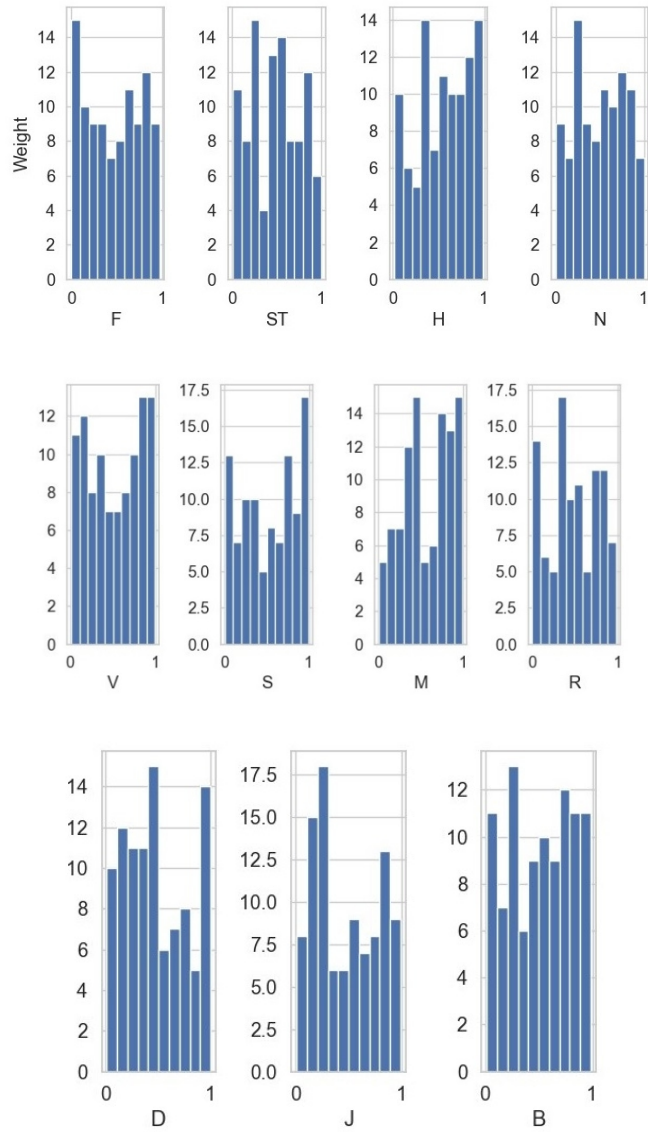

**Figure 3. Histogram plots of all 11 input parameters (sign-symptom).**

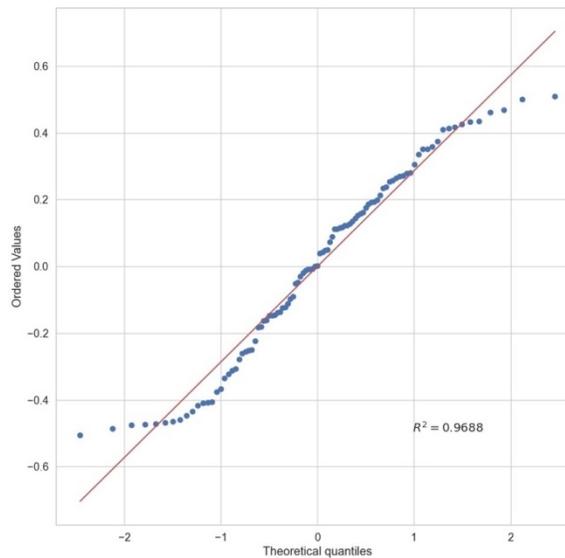

**Figure 4. Probability plot (Q-Q plot) of model residuals.**

$R^2$ calibrating the proportion of variance of the fatality prediction that is predictable from the independent sign-symptoms, showing the goodness of model fit, is shown in **Figure 4**. The $R^2$ from our data scores at *97%* indicating that 97% of the data are distributed close to mean. The result confirms the stability of the VIRDOCD model that is suggestive as the number of data points (100) was not exactly statistically large. From ANOVA, it was also clear that sign-symptoms were largely independent of each other leading to a good statistical fit ($R^2$ = 97%) for the VIRDOCD model.

*Summary of the experimental data analysis:*

1. The database comprised of 100 dengue cases involving 11 input parameters and one output parameter, each with three grades (mild, moderate, and severe). Input parameters were essentially the sign-symptoms of dengue, while the output parameter represents the grade or degree of case fatality.
2. All 11 input and output parameters were assigned weights [0,1] by ten experienced clinicians. Based on their domain knowledge, each case could be represented as an IF-THEN rule and hence, we could consider it not just as a mere database, but rather as a 'rule base' of 100 cases. This 'rule base' is nothing but the domain knowledge or "Clinical eye" of the doctors.
3. From ANOVA it was clear that sign-symptoms jointly and individually influence the grade/severity of case fatality. Each sign-symptom was independent of the other and followed non-Gaussian distributions, as expected. The model showed a good statistical fit ($R^2$ = 97%) and hence, could be used to train the VIRDOCD model.

Coefficient constants, predictors (i.e., sign-symptoms) as in **Table 5**:

**Table 5. Coefficient table**

| Constant | 0 |
|---|---|
| F | 0.0248 |
| ST | -0.1293 |
| H | 0.1161 |
| N | 0.2362 |
| V | -0.0501 |
| S | -0.1125 |
| M | -0.0506 |
| R | -0.0901 |
| D | 0.0397 |
| J | 0.0286 |
| B | -0.1052 |

**B. PERFORMANCE OF VIRDOCD:**

**Table 8**, **Figures 5a** and **5b** represent a sample prediction done by testing VIRDOCD and RF (trained on 75% data) on 25% test cases using 10-fold cross-validation, including error estimation. As discussed, the target of this study is to deliver a Virtual "Clinical Eye", which is nothing but the product of the coefficient values of the weighted sign-symptoms and the added 'bias' value, obtained from equation 5. The coefficient values are the numerical representations of individual 'perception' based judgement. Since no human judgement is completely bias-free, to make VIRDOCD's clinical judgement close to human-judgment, the bias value obtained from equation 5 has been added to the score line. The cumulative scores (Y) for each test case is hence the product of coefficient values (B) and weighted sign-symptoms (X), added with bias ($B_0$).

Table 5 tabulates the Coefficient values for each sign-symptom ($B_0 = 0$) as outlined in Eqn. (5). VIRDOCD output, designated as Calc_Out in Table 6 (column 2), is evaluated by combining the parameters from Table 5 (blue)with remaining 25% of data {F=0.9650, ST=0.3397, H=0.8671, …, J=0.0882, B=0.4717} (green):

***Calc_Out = 0 + 0.0248 x 0.9650 + (-0.1293) x 0.3397 + 0.1161 x 0.8671 + … + 0.0286 x 0.0882 + (-0.1052) x 0.4717 = 0.4562* (moderate outcome),**

where *'Calc_Out'* (2nd column) and *'Target_Out'* (3rd column) for the 1st test case (1st row, denoted by 0) matches real data.

## C. PARAMETRIC STUDY FOR TESTING THE ROBUSTNESS:

Detailed parametric study is an important step to examine the 'individuality' in decision making by VIRDOCD. The study was done in two stages

(i) *Single factor influence*: analyzing system response by individually varying the weights of one of the 11 factors (sign-symptoms) while keeping the other 10 unchanged over a 3-point mMS span: 0.1 (mild), 0.5 (moderate), and 0.9 (severe), and

(ii) *All factor influence*: analyzing multidimensional system response by varying all 11 factors simultaneously over a wider 5-point span: 0.05 (very mild), 0.1 (mild), 0.5 (moderate), 0.9 (severe), 0.95 (very severe).

*Single factor influence (**Table 6**):*

'Nausea (N)' is a specific case in hand. If the other 10 sign-symptoms are restricted to the 'mild' (=0.1) category, the individual N-response (=0.541) records as 'moderate', that is of a higher category. On the other hand, if the other 10 factors are constrained at the 'moderate' level (=0.5), the individual N-response (=0.6355) records as 'severe', that is of the highest category. This is a highly encouraging result as VIRDOCD can be seen to show judgmental independence in analyzing the dengue CFG, akin to that of a team of experienced clinicians. It does not over or under weigh its prediction depending on the initial choice of the mMS category.

### Table 6. Result of Single factor parametric study

|     | 0.1   | 0.5    | 0.9    |
|-----|-------|--------|--------|
| F   | 0.541 | 0.551  | 0.5621 |
| ST  | 0.541 | 0.4893 | 0.4376 |
| H   | 0.541 | 0.5875 | 0.6339 |
| N   | 0.541 | 0.6355 | 0.73   |
| V   | 0.541 | 0.4961 | 0.4511 |
| S   | 0.541 | 0.5208 | 0.5006 |
| M   | 0.541 | 0.505  | 0.4689 |
| R   | 0.541 | 0.5569 | 0.5728 |
| D   | 0.541 | 0.564  | 0.5525 |
| J   | 0.541 | 0.5525 | 0.564  |
| B   | 0.541 | 0.499  | 0.4569 |

*Multi-factor influence (**Table 7**):*
**Table 7** shows how a dengue case is predicted by VIRDOCD when sign-symptom weights are valued at 0.05 (very mild), 0.1 (mild), 0.5 (moderate), 0.9 (severe), 0.95 (very severe). We see that despite all symptoms being weighted as very mild, or mild or severe or very severe, VIRDOCD has retained its own opinion as to the definition of the real 'moderate' throughout, despite varying weights from the clinicians over a much wider range, which again confirms independence in the VIRDOCD prediction profile, as expected with conventional clinicians.

**Table 7. Result of Multi-factor parametric study**

| Weights | 0.05 | 0.1 | 0.5 | 0.9 | 0.95 |
|---|---|---|---|---|---|
| Predicted_Risk | 0.5457 | 0.541 | 0.5641 | 0.4671 | 0.6025 |

*Summary of the performance of VIRDOCD:*

- MLR algorithm has worked well to build the VIRDOCD (Virtual doctor) predictive model.
- Diagnostic accuracy (RMS) – approximately 75%.
- Robust, i.e., not so hypersensitive to the learnt rule base and is able to preserve its individuality.

## D. VALIDATION:

Performance of VIRDOCD has been validated by (i) comparing with another regression-based classifier, such as an RF-based classifier, which has been developed in this study and (ii) comparing with the human clinicians' diagnostic accuracy.

Experimental results show that RF-based classifier shows lesser accuracy (63%), compared to VIRDOCD (75%) (refer to **Table 8** and **Figures 5a** and **5b**), based on the RMS value. While comparing the diagnostic accuracy with human (doctor), study has shown that their overall clinical diagnostic accuracy is about 71.4% (Richens, Lee, & Johri, 2020). Hence, VIRDOCD, for this dataset performs the best.

**Table 8. Prediction of dengue case fatality by VIRDOCD and RF-based classifier and error in prediction, a comparison**

| | Calc_Out | Target_Out | Error | Sq_Error | MSE | RMSE |
|---|---|---|---|---|---|---|
| 0 | 0.456288 | 0.0447 | -0.4116 | 0.1694 | | |
| 1 | 0.686331 | 0.9845 | 0.2982 | 0.0889 | | |
| 2 | 0.638174 | 0.8304 | 0.1922 | 0.0369 | | |
| 3 | 0.46606 | 0.6325 | 0.1664 | 0.0277 | | |
| 4 | 0.496371 | 0.9212 | 0.4249 | 0.1805 | | |
| 5 | 0.434475 | 0.5008 | 0.0663 | 0.0044 | | |
| 6 | 0.499702 | 0.0453 | -0.4544 | 0.2065 | | |
| 7 | 0.667975 | 0.8210 | 0.1530 | 0.0234 | | |
| 8 | 0.390643 | 0.1966 | -0.1940 | 0.0376 | | |
| 9 | 0.459032 | 0.1884 | -0.2706 | 0.0732 | | |
| 10 | 0.454379 | 0.4701 | 0.0157 | 0.0002 | | |
| 11 | 0.405012 | 0.3161 | -0.0889 | 0.0079 | 0.0623 | 0.2497 |
| 12 | 0.377801 | 0.7712 | 0.3934 | 0.1547 | | |
| 13 | 0.591266 | 0.3732 | -0.2181 | 0.0476 | | |
| 14 | 0.494619 | 0.7657 | 0.2710 | 0.0735 | | |
| 15 | 0.508836 | 0.6705 | 0.1617 | 0.0261 | | |
| 16 | 0.433667 | 0.4949 | 0.0612 | 0.0038 | | |
| 17 | 0.369092 | 0.2233 | -0.1458 | 0.0213 | | |
| 18 | 0.462861 | 0.4012 | -0.0617 | 0.0038 | | |
| 19 | 0.375639 | 0.1089 | -0.2668 | 0.0712 | | |
| 20 | 0.48632 | 0.5401 | 0.0538 | 0.0029 | | |
| 21 | 0.616385 | 0.2523 | -0.3640 | 0.1325 | | |
| 22 | 0.403173 | 0.0533 | -0.3499 | 0.1224 | | |
| 23 | 0.389266 | 0.2202 | -0.1691 | 0.0286 | | |
| 24 | 0.530327 | 0.4149 | -0.1154 | 0.0133 | | |

| | Calc_Out | Target_Out | Error | Sq_Error | MSE | RMSE |
|---|---|---|---|---|---|---|
| 0 | 0.98 | 0.0447 | -0.9353 | 0.8748 | | |
| 1 | 0.686331 | 0.9845 | 0.2982 | 0.0889 | | |
| 2 | 0.738174 | 0.8304 | 0.0922 | 0.0085 | | |
| 3 | 0.660597 | 0.6325 | -0.0281 | 0.0008 | | |
| 4 | 0.896371 | 0.9212 | 0.0249 | 0.0006 | | |
| 5 | 0.634475 | 0.5008 | -0.1337 | 0.0179 | | |
| 6 | 0.099702 | 0.0453 | -0.0544 | 0.0030 | | |
| 7 | 0.667975 | 0.8210 | 0.1530 | 0.0234 | | |
| 8 | 0.191643 | 0.1966 | 0.0050 | 0.0000 | | |
| 9 | 0.119032 | 0.1884 | 0.0694 | 0.0048 | | |
| 10 | 0.454379 | 0.4701 | 0.0157 | 0.0002 | | |
| 11 | 0.315012 | 0.3161 | 0.0011 | 0.0000 | 0.1356 | 0.3683 |
| 12 | 0.377801 | 0.7712 | 0.3934 | 0.1547 | | |
| 13 | 0.341266 | 0.3732 | 0.0319 | 0.0010 | | |
| 14 | 0.494619 | 0.7657 | 0.2710 | 0.0735 | | |
| 15 | 0.528836 | 0.6705 | 0.1417 | 0.0201 | | |
| 16 | 0.936674 | 0.4949 | -0.4418 | 0.1952 | | |
| 17 | 0.869092 | 0.2233 | -0.6458 | 0.4171 | | |
| 18 | 0.402861 | 0.4012 | -0.0017 | 0.0000 | | |
| 19 | 0.175639 | 0.1089 | -0.0668 | 0.0045 | | |
| 20 | 0.38632 | 0.5401 | 0.1538 | 0.0236 | | |
| 21 | 0.916385 | 0.2523 | -0.6640 | 0.4410 | | |
| 22 | 0.813173 | 0.0533 | -0.7599 | 0.5775 | | |
| 23 | 0.892657 | 0.2202 | -0.6725 | 0.4522 | | |
| 24 | 0.330327 | 0.4149 | 0.0846 | 0.0072 | | |

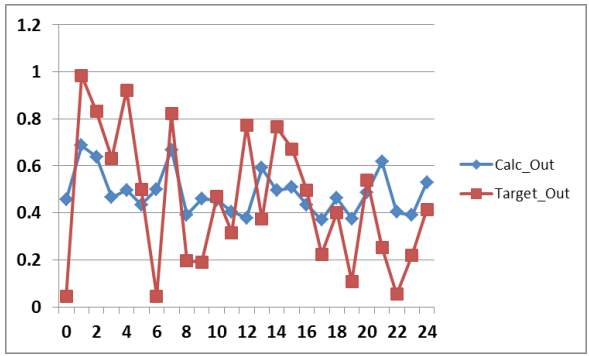

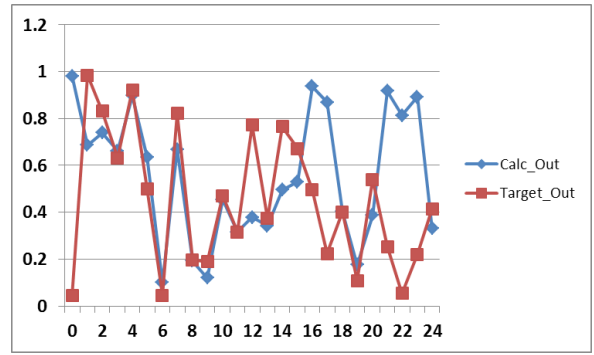

**Figure 5a. Calculated vs Actual outputs**     **Figure 5b. Calculated vs Actual outputs (RF)**

**Figure 5: Comparison between model versus real data**

**IV. DISCUSSION:**

Both as a diagnostic tool and also as a medical aid, digital healthcare is under the radar. The last two decades have seen increasing implementation of Statistical Machine Learning (SMI) tools to assist clinical practices in screening, diagnosing and grading an illness. However, no study has utilized clinicians' rule-based 'learning' while arriving at a prediction by analyzing the weighted sign-symptoms to conclude on the probability of a disease, or its severity grade. In other words, no work has been reported that has attempted to quantify and then translate the 'clinical eye' of a human physician into data-validated diagnosis. The analytics within the proposed tool (VIRDOCD) is based on experience-based weighing of the sign-symptoms, then combining them towards a cumulative outcome as a probabilistic conclusion, that we call as the 'medical diagnosis' (Chattopadhay, 2013). The follow-up therapeutic routine is based on the correctness of the diagnosis. For more complicated cases involving a team of clinicians in a medical board, the specialists first independently analyze the sign-symptoms and then take an arithmetic mean across the board to converge to a unified opinion. Bulk of the relevant literature analyze patient data based on their arbitrary choice of classifiers without attempting to causally link their algorithms with deductive reasoning from the human clinicians. There is also the curse of subjectivity. Classifiers failing with a certain dataset are not necessarily crippled against all datasets, and vice versa. These limitations has traditionally coerced against realistic implementation of computational diagnostics in a real medical setup.

The present study is an attempt to develop an SMI-based tool (VIRDOCD) that applies the rule base of the human clinicians (weighted sign-symptoms and possible grade of the illness) and hence can better serve as a diagnostic aid to clinicians in screening and grading Dengue cases. VIRDOCD is structured on recursive MLR that requires training the algorithm with weighted patient data (assigned by a group of doctors), and then combining the individual diagnostics to identify a human error-free diagnostic decision. Table 1 shows the sign-symptoms of a cluster of 100 dengue infected (at various levels) patients that are initially trained on 80% data, then performance tested against human judgments. Robustness in the decision making process is a key target that we measure by examining whether the final outcome from VIRDOCD suffers from judgmental bias introduced by the training rules, or are genuinely neutral. The agreement with predictions from a specialist medical board (10 senior clinicians) confirms that the algorithm works and is actually more accurate. Its performance is further validated against another regression-based classifier, developed using RF algorithm, and found to be superior to it, as well as against the overall predicting accuracy of human clinicians.

Although MLR is apparently a (computationally) 'hard' technique, yet, the coefficient values (refer to equation 5 and Table 5) vary across the weighted datasets that are influenced by the statistical properties of the data, such as its distribution, pattern, interconnectedness amongst the attributes (ANOVA, refer to Table 4), and statistical significance (p-values with CI 95%). It is important to note that the MLR algorithm in VIRDOCD has been developed relying on the inherent malleability of the coefficient values. Therefore, the algorithmic 'hardness' has been effectively reduced. Coefficient values derived from VIRDOCD, therefore, can clone human-like perception rather than mechanized numbers. Finally, VIRDOCD's robustness has been tested through parametric studies, where some parameters weights were kept constant varying others. To our satisfaction, VIRDOCD was able to retain its judgmental individuality (refer to Table 7 & 8).

## V. CONCLUSIONS AND FUTURE WORK:

Statistical methods such as ANOVA and MLR are useful techniques to develop a predictive epidemic model. ANOVA provides insights into the data structure by analyzing the 'intra' and 'inter'-group variations, data distribution, and effect of each predictor (sign-symptoms) on disease outcome (fatality grade/severity). On the other hand, MLR-based predictive modeling throws light on how 'fit' the model is, and how each predictor influences outcome prediction. VIRDOCD (virtual doctor) is an MLR-based predictive model (a new doctor) that is able to learn through the rule base (Clinical eyes) of human doctors, develop its own 'understanding' (by the Coefficient values obtained through MLR process). and finally develop its own Clinical eye (product of its understanding and the rule base given by the human doctors). Thus it combines the best of both worlds. Comparing the diagnostic accuracy of VIRDOCD (ca 75%) with another regression-based classifier (RF, ca 63%) and human clinical diagnoses at 71.4% (Richens, Lee, & Johri, 2020), the measure seems well balanced and trained for accurate prediction. Another important feature of VIRDOCD is its ability to retain its non-biased attribute in decision making, unlike human doctors who are likely to show subjective fluctuations in their patient evaluation.

Work is ongoing to introduce a chromatic RGB-styled sign-symptom grading, which can vary continuously between 0 and 1 to make medical predictions objectively subjective. We are in the process of introducing 'stochastic assessment kernels' before and after VIRDOCD data training that could then distribute numbers within these intermediate regimes generalizing the 3-point scale to a higher dimensional construct.

**CONFLICTS OF INTEREST:**

None

**DATA AVAILABILITY STATEMENT:**

The data used in this work have been collected and curated by a team of clinicians, including Dr S Chattopadhyay. These are real human data and hence confidential that cannot be shared even anonymously due to hospital data embargo. All codes used in this work have been personally written by Dr S Chattopadhyay using Python 3.8.3 and Spyder editor version 4.1.5 in Windows 10 Pro 64 bits OS, Processor – Intel (R) Core (TM) i5 3360M CPU@ 2.80GHz, 8GB RAM. Data presented in plots may be personally shared on request.